# Fundamental Performance Limits of Carbon Nanotube Thin-Film Transistors Achieved Using Hybrid Molecular Dielectrics

Vinod K. Sangwan[1], Rocio Ponce Ortiz[2], Justice M. P. Alaboson[1], Jonathan D. Emery[1], Michael J. Bedzyk[1,3], Lincoln J. Lauhon[1], Tobin J. Marks[1,2,*], Mark C. Hersam[1,2,4,*]

[1]Department of Materials Science and Engineering, Northwestern University, Evanston, IL 60208, USA

[2]Department of Chemistry, Northwestern University, Evanston, IL 60208, USA

[3]Department of Physics and Astronomy, Northwestern University, Evanston, IL 60208, USA

[4]Department of Medicine, Northwestern University, Evanston, IL 60208, USA

ABSTRACT:

In the past decade, semiconducting carbon nanotube thin films have been recognized as contending materials for wide-ranging applications in electronics, energy, and sensing. In particular, improvements in large-area flexible electronics have been achieved through independent advances in post-growth processing to resolve metallic *versus* semiconducting carbon nanotube heterogeneity, in improved gate dielectrics, and in self-assembly processes. Moreover, controlled tuning of specific device components has afforded fundamental probes of the trade-offs between materials properties and device performance metrics. Nevertheless, carbon nanotube transistor performance suitable for real-world applications awaits understanding-based progress in the integration of independently pioneered device components. We achieve this here by integrating high-purity semiconducting carbon nanotube films with a custom-designed hybrid inorganic-organic gate dielectric. This synergistic combination of materials circumvents conventional design trade-offs, resulting in concurrent advances in several transistor performance metrics such as transconductance (6.5 µS/µm), intrinsic field-effect mobility (147 cm$^2$/Vs), sub-threshold swing (150 mV/decade), and on/off ratio (5 x 10$^5$), while also achieving hysteresis-free operation in ambient conditions.

TEXT:

Carbon nanotube (CNT) thin films[1] are promising semiconductors for diverse applications including large-area printed electronics,[2-7] high-frequency devices,[8,9] and light-emitting diodes.[10,11] However, thin-film transistors (TFTs) fabricated from as-grown heterogeneous CNT films are intrinsically limited in performance due to contamination by metallic nanotubes.[12-14] Furthermore, device performance is strongly constrained by gate dielectric details, including capacitive coupling to the channel, interfacial scattering, and trapped charges. To date, research efforts have largely focused on independent issues such as CNT purity,[15-17] CNT density,[5,18,19] channel geometry,[2,20] and gate-dielectric properties[3,21,22] to improve specific device metrics, often at the expense of others. More attractive for the ultimate incorporation of CNT films in low-power, large-area electronics is a more holistic approach whereby the gate dielectric and semiconductor channel are approached synergistically, resulting in simultaneous optimization of multiple device metrics, including important but less-discussed ones such as hysteresis.

The recent demonstration of scalable methods for producing monodisperse semiconducting CNTs[15-17] offers immediate attractions and has afforded enhanced device performance such as large on/off ratios,[6] large channel conductance,[3,20] high field-effect mobilities,[7,17,18] together with large-area uniformity and yield.[3,6,7] Moreover, minimization of metallic CNT content (<1%) permits use of thicker CNT films which can sustain higher current densities in the on-state without significantly increasing the off-state current. While these thick CNT films can achieve higher current densities, the on/off ratios when using low-capacitance dielectrics are compromised due to CNT-CNT screening. Imperfections in the gate dielectric and/or suboptimal dielectric-CNT interfaces also degrade other key performance metrics such as hysteresis, threshold voltage, and sub-threshold swing – properties that must be concurrently improved for effective implementation of low-power, high-speed CNT TFT-based electronics.

Toward this end, we report here the integration of >99% pure semiconducting CNTs with a new class of nanoscopic high-capacitance (630 nF/cm$^2$) hybrid inorganic-organic gate dielectrics[23] to achieve TFT performance unconstrained by traditional trade-offs. The resulting devices simultaneously exhibit low operating voltages (4 V), low sub-threshold swings (150 mV/decade), high normalized on-state conductance (8.5 µS/µm), high transconductance (6.5 µS/µm), and high intrinsic field-effect mobilities (147 cm$^2$/Vs) with high on/off ratios (5 x 10$^5$) in ambient conditions. This unique combination of hybrid gate dielectrics with monodisperse semiconducting CNTs is compatible with low-temperature, large-area processing, thus offering applications in low-power TFT-based electronics. These devices also exhibit negligible hysteresis in transfer characteristics, unlike those fabricated with conventional oxide dielectrics, and avoid the ambipolarity that increases power consumption for CNT TFT circuits based on gel dielectrics.[4] The hybrid dielectric ("VA-SAND"), fabricated by combining inorganic atomic layer deposition (ALD) with vapor phase organic self-assembly, can be grown with precise thickness control and combines layers of π-conjugated donor-acceptor building blocks, self-assembled *via* hydrogen bonding (κ ~ 9),[23,24] with ultra-thin (~2 nm) layers of ALD-derived Al$_2$O$_3$ to enhance stability and dielectric characteristics (Fig. 1a).

**Gate dielectric fabrication and characterization**

VA-SAND was grown on degenerately doped Si/SiO$_2$ substrates containing 1.8 nm thick native oxide (see Methods for details). VA-SAND microstructure and morphology were characterized by X-ray reflectivity (XRR) and AFM, while leakage current and capacitance-voltage (C-V) analysis were carried out on metal-insulator-semiconductor (MIS) capacitors. To highlight the differences between VA-SAND and purely inorganic oxide dielectrics grown by ALD, MIS capacitors and TFTs fabricated on ALD-grown $Al_2O_3$ (6-AO) of the same total thickness as VA-SAND (6 nm) were also characterized. The background-subtracted XRR data and model fits for VA-SAND on $SiO_x$ are shown in Fig. 1b. The electron density profile (normalized to the electron density of Si) resulting from the best fit of the XRR data using a 4-slab density model is plotted as a function of distance from the surface of the native oxide in Fig. 1c.[25] The layer thicknesses are derived from the inflection points in the electron density profiles and are interpreted as the boundaries between layers.[26] The extracted thickness of the native oxide, the $Al_2O_3$ underlayer, the organic layer, and the $Al_2O_3$ capping-layer are 1.8 nm, 1.5 nm, 2.5 nm, and 2.0 nm, respectively. AFM images of VA-SAND (inset in Fig. 1b) reveal an RMS roughness of 0.65 nm, in agreement with the RMS roughness of 0.7 nm extracted from the XRR analysis. Both VA-SAND and 6-AO exhibit comparable leakage current densities of $10^{-7}$ $A/cm^2$, up to 7 orders of magnitude lower than that of the $SiO_2$ native oxide as the top-electrode bias is varied from –2 V to 2 V (Fig. 1d). Note that VA-SAND exhibits 37% higher capacitance (630 $nF/cm^2$) than 6-AO (460 $nF/cm^2$) with the substrate in accumulation (V > 1.2 V) due to the higher κ of the organic layer (Fig. 1e). The capacitance decreases as the bias is varied from 1.2 V to -0.5 V due to the formation of the depletion region in the Si substrate and becomes constant for V < $V_{th}$ = -0.5 V. VA-SAND also exhibits lower current leakage ($10^{-7}$ $A/cm^2$) and higher capacitance (630 $nF/cm^2$) than previously reported vapor-deposited V-SAND[24] due to the reduced thickness combined with the higher-κ of the robust upper inorganic layer (*vide infra*).

The dielectric constants of individual layers were determined by parallel plate capacitor analysis (see Methods) and were found to be 3.9, 6.0, 9.5, and 8.0 for the native oxide, the underlayer, the organic layer, and the upper capping-layer, respectively. This thickness and dielectric constant analysis is consistent with the following observations: 1) The lower electron density (higher dielectric constant) of the capping layer *versus* the underlayer (Fig. 1c); 2) The lower thickness of the organic layer (2.5 nm) compared to the length of two head-to-tail hydrogen-bonded chromophore molecules (3.4 nm);[24] 3) The thicker upper capping-layer (2 nm) *versus* that of the underlayer (1.5 nm) with the same number of ALD growth cycles; and 4) A rougher capping layer-organic layer interface *versus* ALD-grown $Al_2O_3$. These observations suggest significant intermixing of the chromophore and capping $Al_2O_3$ layers yielding an effective dielectric constant of 8 for the intermixed capping layer that is in between 9.5 for the organic layer and 6.0 for $Al_2O_3$ (Fig. 1(a)). Thus, intermixing explains the lower electron density, higher dielectric constant, and increased thickness and roughness of the capping layer. The net effective dielectric constant of a 6 nm thick VA-SAND layer (7.8 nm thick, including the native oxide) is found to be $\kappa_{\text{VA-SAND}}$ = 6.36 (5.55). The effective oxide thickness (EOT) of VA-SAND without (with) the native oxide is determined to be 3.68 nm (5.48 nm).

**Carbon nanotube thin-film transistor fabrication, characterization, and analysis**

Fig. 2a shows an optical micrograph of the semiconducting single-walled CNT band in a centrifuge tube after two iterations of density gradient ultracentrifugation (DGU) of arc-discharge-derived single-walled CNTs (Supplementary Section 1). The relative content of semiconducting CNTs is calculated to be 99% by comparing the relative area under the metallic and semiconducting peaks in the optical absorbance spectra, Fig. 2b.[15,17] Semiconducting CNT enrichment is clearly evident in the decreased (increased) metallic M11 (semiconducting S22 and S33) peaks in the sorted CNT solution compared to the as-grown CNTs. Bottom-contact CNT TFTs were next fabricated on VA-SAND using photolithography (Methods) with channel lengths $L$ varying from 5 µm to 50 µm and channel width $W$ = 100 µm (Figs. 2c, 2d, 2e). The CNT average length was determined to be 1.36 ± 0.92 µm from AFM analysis of a large ensemble of 334 nanotubes (Supplementary Section 2).

The origin of the resulting exceptional CNT/VA-SAND TFT performance is illustrated through a systematic analysis of device metrics at four different CNT film densities: density-1 = 5.5 ± 0.9 CNTs/µm$^2$; density-2 = 13.3 ± 1.7 CNTs/µm$^2$; density-3 = 22.7 ± 1.9 CNTs/µm$^2$; and density-4 = 27.1 ± 2.5 CNTs/µm$^2$ (Fig. 2f – 2i). These CNT TFTs exhibit p-type behavior in ambient at low biases of $V_g$ = -2 V to 2 V, and $V_d$ = 0.5 V to -2 V (Fig. 3a). The negligible hysteresis of these TFTs on VA-SAND compared to that on 6-AO (Fig. 3a and 3b) suggests significantly lower VA-SAND trap charge densities and/or favorably modified surface properties compared to the conventional oxide ALD dielectric of the same thickness. Note that the present CNT/VA-SAND devices exhibit small threshold voltages (< 1V) and ultra-low sub-threshold slopes, as low as ~100 mV/decade, compared to the quantum limit of ~70 mV/decade at room temperature (Supporting Fig. S4b), making these TFTs suitable candidates for low-power, high-speed circuits. Note that the sub-threshold slope as well as off-current ($I_{off}$, within the noise level of the instrumentation ~10 pA) remain relatively independent of drain bias (Fig. 3c).

We next assess device parameters that underlie the performance of digital circuit building blocks such as inverters and ring-oscillators.[12,13] First, high field-effect mobility (high transconductance) is necessary to achieve large voltage gain inverters in high speed circuits. Second, a low-operating voltage and high on/off ratio (i.e., low off-current) is necessary to minimize power dissipation. Finally, a reduced channel area (i.e., high current-capacity or normalized conductance) is desired to minimize parasitic capacitance in high-frequency digital circuitry. Note that while individual CNTs have large current-carrying capacities and high field-effect mobilities,[27] the effective field-effect mobility of CNT films is significantly reduced due to additional resistance from CNT-CNT junctions. Further reduction in the *estimated mobility* can result from assumptions made about the morphology of percolating CNT films in calculating the gate capacitance. There are two methods commonly used to estimate the capacitance of a random network CNT film. In the first, the CNT film is assumed to be continuous in a parallel plate geometry, affording a capacitance $C_g = C_{pp} = \frac{\varepsilon_{ox}}{t_{ox}}$ (630 nF/cm$^2$), where $\varepsilon_{ox}$ and $t_{ox}$ are the dielectric constant and thickness of the gate-dielectric, respectively. Note that the assumed gate capacitance of 630 nF/cm$^2$ is the upper limit of the capacitance of VA-SAND at the onset of the inversion region (V$_{top\ electrode}$ = 1 V for an MIS capacitor on n-type Si (Fig. 1e) and $V_g$ = -1 V for CNT TFTs). Thus, the reported field-effect mobilities actually underestimate the actual values. The second method takes into account electrostatic coupling between CNTs as well as the quantum capacitance of CNTs[2,28] to obtain the intrinsic capacitance of the CNT films, $C_{IN}$. The dependence of $C_{IN}$ on CNT density and gate dielectric capacitance is illustrated in Supporting Section 4. Overestimation of capacitance in $C_{PP}$ is more critical in the case of sparse CNT networks and high capacitance gate dielectrics. For completeness, we report both the parallel-plate field-effect mobility ($\mu_{PP}$) and intrinsic field-effect mobility ($\mu_{IN}$) calculated from $C_{PP}$ and $C_{IN}$, respectively, using $\mu = \frac{L}{C_g V_d W} \frac{\partial I_d}{\partial V_g}$, where $I_d$, $V_d$ and $V_g$ are drain current, drain voltage, and gate voltage, respectively.

Figs. 3d and 3e show the output characteristics of a low density (density-1) and a high density (density-4) CNT/VA-SAND TFT ($L$ = 5 μm and $W$ = 100 μm), and Fig. 4a compares the transfer characteristics of four different CNT density TFTs having the same channel dimensions. The drain current is varied over two orders of magnitude to determine the CNT density for optimum device performance. At all densities, the CNT TFTs show linear behavior at low $V_d$, suggesting ohmic CNT-electrode contacts. Since width-normalized on-state conductance ($G/W = I_d/(V_d \cdot W)$) is a commonly used figure-of-merit for current-carrying capacity, both $G/W$ and $I_d$ (at $V_d$ = -100 mV) are plotted in linear and semi-log plots, respectively (Fig. 4a). The lowest CNT density (density-1), 5.5 CNTs/ μm$^2$, is above the percolation threshold[29] $\rho_{th} = \frac{4.24^2}{\pi L_{CNT}^2} =$ 3.09 CNTs/μm$^2$ (average CNT length $L_{CNT}$ = 1.36 μm), while the highest CNT density (density-4), 27.1 CNTs/μm$^2$, exhibits a low on-state sheet resistance (at $V_g$ = -2 V) of 16.8 kΩ/square. The present CNT TFTs show dominantly p-type behavior with gradually increasing ambipolarity and larger $I_{off}$ ($I_{off}$ = minimum $I_d$) at higher CNT densities. Ambipolar behavior in thicker CNT films may reflect band-to-band tunneling due to increased fractions of small diameter CNTs[30] and/or decreased interaction of CNTs with adsorbates in thicker films.[31]

The effect of CNT density on device performance is illustrated in Fig. 4b where the relevant device parameters (averaged over 5 devices) are plotted as a function of CNT density. The average width-normalized on-current ($I_{on}/W$, $I_{on} = I_d$ at $V_g$ = -2 V) as well as average off-current of the devices ($W$ = 100 μm) increases with CNT density. $I_{on}/W$ increases by two orders of magnitude from density-1 to density-3 films and then increases only marginally (~30%) for density-4 CNT films. In contrast, $I_{off}$ increases by only an order of magnitude for the first 3 CNT densities, but increases by more than 2 orders of magnitude for density-4 CNTs. These currents imply an almost constant on/off ratio up to density-3 and then more than two orders of magnitude decreased on/off ratio for density-4. In the lower part of Fig. 4b, the sub-threshold slope (SS) and drain voltage- and width-normalized transconductance ($g_{m,nor} = \frac{1}{V_d W} \frac{\partial I_d}{\partial V_g}$ in the linear regime, -2 V < $V_g$ < -1 V) are plotted as a function of CNT density. Note that the sub-threshold slope remains close to 150 mV/decade up to density-3 films and then increases to 450 mV/decade for density-4 films. The normalized transconductance increases by 50x from density-1 to density-3 and then increases by only 20% for density-4. The trade-off between on/off ratio and field-effect mobility indicated in Fig. 4c reveals increasing $\mu$ up to density-3 without degradation in the on/off ratio. Further increases in CNT density result in decreased on/off ratio without a significant increase in field-effect mobility. This trade-off may reflect the significant role played by the low fraction of metallic CNTs and/or the effects of CNT-CNT screening in thick monodisperse CNT films. Thus, integration of the hybrid VA-SAND gate dielectric with high-purity thick monodisperse CNTs allows optimization of device performance (density-3) to an average field-effect mobility of $\mu_{PP}$ = 42 cm$^2$/Vs and $\mu_{IN}$ = 136 cm$^2$/Vs at an average on/off ratio ~10$^5$. In contrast, as-grown CNTs produce low on/off ratios at significantly lower coverages due to the lower percolation threshold from the large fraction (30%) of long (~10 μm) metallic CNTs,[13,19] whereas monodisperse CNTs on low capacitance gate-dielectrics (e.g., 300 nm SiO$_2$) exhibit low on/off ratios, likely due to the onset of CNT-CNT screening at lower CNT densities.[18]

Density-3 CNT TFTs were further characterized to investigate large-area uniformity and channel geometry effects. The average transfer characteristics of 7 density-3 CNT TFTs spread over ~2 mm in Fig. 4d reveal excellent device-to-device uniformity. On-currents remain within ±18% and on/off ratios within one order of magnitude of the respective average values. Uniformity in such devices reflects the self-limiting thin-film growth mechanism of vacuum filtration.[14] Note that a consistent TFT threshold voltage (within 100 mV of -0.5 V) is highly desirable for large-area low-voltage CNT circuitry. Fig. 4g shows normalized on-state conductance ($G_{on}/W = G/W$ at $V_g$ = -2 V), intrinsic field-effect mobility ($\mu_{IN}$), and on/off ratio of density-3 CNT TFTs with $L$ varying from 5 µm to 50 µm ($W$ = 100 µm). $G_{on}/W$ decreases sharply with $L$ while the on/off ratio increases slightly with $L$. A slight fall in $\mu_{IN}$ with $L$ can be attributed to the sub-linear length dependence of resistivity in percolating CNT networks, in agreement with previous reports.[19,32]

We next examine the principal performance parameters of CNT/VA-SAND TFTs in the context of previously reported CNT TFT design trade-off relationships.[2,3,5-7,15,18-21,33-39] Figs. 5a-e show on/off ratios plotted as a function of normalized on-state conductance ($G_{on}/W$), normalized transconductance ($g_{m,nor}$), operating voltage, parallel-plate field-effect mobility ($\mu_{PP}$), and intrinsic field-effect mobility ($\mu_{IN}$), respectively. A common legend for all the plots is shown in Fig. 5f. Note that the transconductance data from the literature are also normalized with respect to the reported channel widths and drain biases. The best available performance parameters were extracted from the literature on random CVD-grown CNTs and solution-processed and purified semiconducting CNT TFTs, and are then contrasted with optimized density-3 CNT/VA-SAND TFT data. Figures 5a-c show data for 7 short channel length devices ($L$ = 5 µm, $W$ = 100 µm) taken from the transfer plots in Fig. 4d. Figs. 5d,e show data from all density-3 CNT TFTs, including devices with longer channel lengths from the transfer plots of Fig. 4e. Although as-grown CNTs provide larger normalized conductance than monodisperse CNTs of the same network density, the metallic CNTs in heterogeneous mixtures significantly erode on/off ratios.[19,39] Note also that reduced on/off ratios result from high densities of monodisperse CNTs due to increased CNT-CNT screening.[6,18] The present CNT films afford the highest normalized on-state conductance at an on/off ratio = $10^6$ reported to date for CNT TFTs. The present random CNT/VA-SAND TFTs show larger on-state conductance at higher on/off ratios than ambipolar TFTs using thick CNT films on high-capacitance ion-gel dielectrics[3] (~10 µF/cm$^2$) and top-gated ambipolar TFTs having aligned monodisperse CNT strips.[20] P-type CNT TFTs show higher on/off ratios than ambipolar devices where both minority electrons and holes are present in the channel in the off-state, resulting in larger off-currents.[40] The present devices also exhibit significantly higher transconductance at an on/off ratio ~$10^5$ due to the combined high conductance and low voltage operation (Fig. 5c). As expected, there are no obvious trends in on/off ratio *versus* operating voltage in previously reported devices due to the large variations in operation strategies, device geometries, and gate dielectric materials.

The trade-off between on/off ratio and field-effect mobility (Figs. 5d,e) is similar to that of normalized on-state conductance and transconductance with larger (smaller) mobilities at lower (higher) on/off ratios. The present devices outperform the majority of the literature devices ($\mu_{PP}$ = 45 cm$^2$/Vs; $\mu_{IN}$ = 147 cm$^2$/Vs at on/off ratio of 5 x $10^5$) with comparable $\mu_{PP}$ and lower $\mu_{IN}$ than high-quality CVD-grown CNT TFTs.[5] Such devices exhibit high intrinsic mobility (~650 cm$^2$/Vs) due to the reduced intrinsic gate capacitance ($C_{in}$, Supporting Section 4) of a very sparse network of highly conductive long CNTs with possible covalent bonds at CNT-CNT junctions. Note however that the present devices show 100x higher on-state conductance and transconductance and 5x lower operating voltage with significantly reduced hysteresis and sub-threshold swing.

**Conclusions**

In summary, we have approached the fundamental performance limits for 99% purity semiconducting CNTs *via* integration with a high-capacitance hybrid inorganic-organic gate dielectric. Since the VA-SAND gate capacitance (630 nF/cm$^2$) approaches the quantum capacitance of CNT films (~1 μF/cm$^2$ for the density-3 CNT TFTs), further increases in gate capacitance may not yield significantly enhanced performance. Note that the performance reported here for CNT/VA-SAND TFTs compares favorably with devices fabricated from competing semiconducting materials such as polycrystalline Si,[41] organics,[42] and other inorganics.[43] The attractions of monodisperse semiconducting CNT inks include excellent compatibility with ink-jet printing,[3] mechanical flexibility, and environmental stability, making them promising candidates for next-generation printed electronics.

**Methods**

**VA-SAND growth and characterization.** Twenty cycles of ALD-derived Al$_2$O$_3$ using trimethylaluminum and water as precursors were first grown on heavily doped n-Si (100) substrates at 100 °C (Savannah, Cambridge NanoTech), followed by thermal evaporative deposition of the V-SAND organic layer under high vacuum (10$^{-6}$ Torr) at 25 °C. Growth was carried out as described earlier,[21] i.e., at 0.1 - 0.2 Å/sec to obtain a 3.4 nm thick bilayer of two head-to-tail hydrogen-bonded π-molecules. Finally, the organic layer was capped with an additional 20 cycles of an ALD-derived Al$_2$O$_3$ protective layer at 100 °C. MIS capacitors were fabricated by thermal evaporation of 50 nm thick Au electrodes onto the dielectric layers through shadow masks. Leakage I-V measurements were carried out in ambient using a femto-amp Keithley source-meter, and C-V measurements were made at 10 kHz using an HP 4192A impedance analyzer. The capacitances of VA-SAND and 6-AO are modeled as four and two parallel plate capacitors in series, respectively.

$$\frac{1}{C_{VA-SAND}} = \frac{1}{C_{native-oxide}} + \frac{1}{C_{underlayer}} + \frac{1}{C_{chromophore}} + \frac{1}{C_{capping}}$$

$$\frac{1}{C_{6-AO}} = \frac{1}{C_{native-oxide}} + \frac{1}{C_{Al_2O_3}}$$

$$C_i = \frac{k_i \varepsilon_0}{d_i}$$

where, $C_i$, $\kappa_i$, $d_i$ and $\varepsilon_0$ are capacitance per unit area, dielectric constant, thickness of the dielectric layer, and permittivity of free space, respectively. The thickness of native oxide is found to be 1.8 nm by ellipsometery (J.A. Woolam Co. M2000V VASE) on a blank substrate. The dielectric constant of ALD-grown $Al_2O_3$ was independently determined to be 6.0 from an MIS capacitor fabricated on 6 nm thick $Al_2O_3$ grown on a chemically etched Si substrate (without native oxide). The dielectric constant of the organic layer was previously determined as 9.5 by experiment and theoretical modeling.[24]

**X-ray reflectivity characterization of VA-SAND.** XRR data was acquired using an 18 kW Rigaku ATXG diffractometer equipped with a Cu rotating anode (λ = 1.541 Å) equipped with a NaI scintillation detector. X-rays were conditioned with a multilayer parabolic mirror and collimated to 5.0 mm x 0.1 mm (height × width), yielding an incident beam flux of ~1 × 10$^8$ at the sample surface.

**CNT TFT fabrication.** Purification of 99% semiconducting arc-discharged CNTs was achieved by two iterations of DGU as described in detail in Supplementary Section 1.[15-17] CNT films with four different network densities were prepared by vacuum filtration of 30 μL, 60 μL, 90 μL, and 150 μL of 99% semiconducting CNT suspensions (diluted with 2 mL 1% SC:DI $H_2O$) onto 1.42 cm² mixed cellulose ester membranes (Millipore, pore size = 50 nm). The self-limiting vacuum filtration process results in a uniform coating of the CNT film. The CNT films were then rinsed with 100 mL DI $H_2O$ to remove residual surfactant. Vacuum filtration affords large area, clean, and uniform CNT films with excellent control over network density. The TFT source-drain electrodes (Cr/Au: 2/50 nm) were defined on VA-SAND by photolithography, thermal evaporation of the metals, and a lift-off process. An additional 15 nm thick $Al_2O_3$ film was grown by ALD on patterned photoresist before electrode metallization to achieve robust electrical probing in a Cascade Microtech probe station (schematic in Fig. 2d). CNT films were transferred onto patterned source-drain electrodes by dissolving the filter membranes in acetone vapor.[17] The CNT films were then annealed in air at 225 °C for 1 h to further remove residual impurities. Finally, bottom-gate bottom-contact CNT TFT channels were defined using photolithography and reactive ion etching to obtain a channel width ($W$) of 100 μm and channel lengths ($L$) varying from 5 μm to 50 μm.

FIGURES

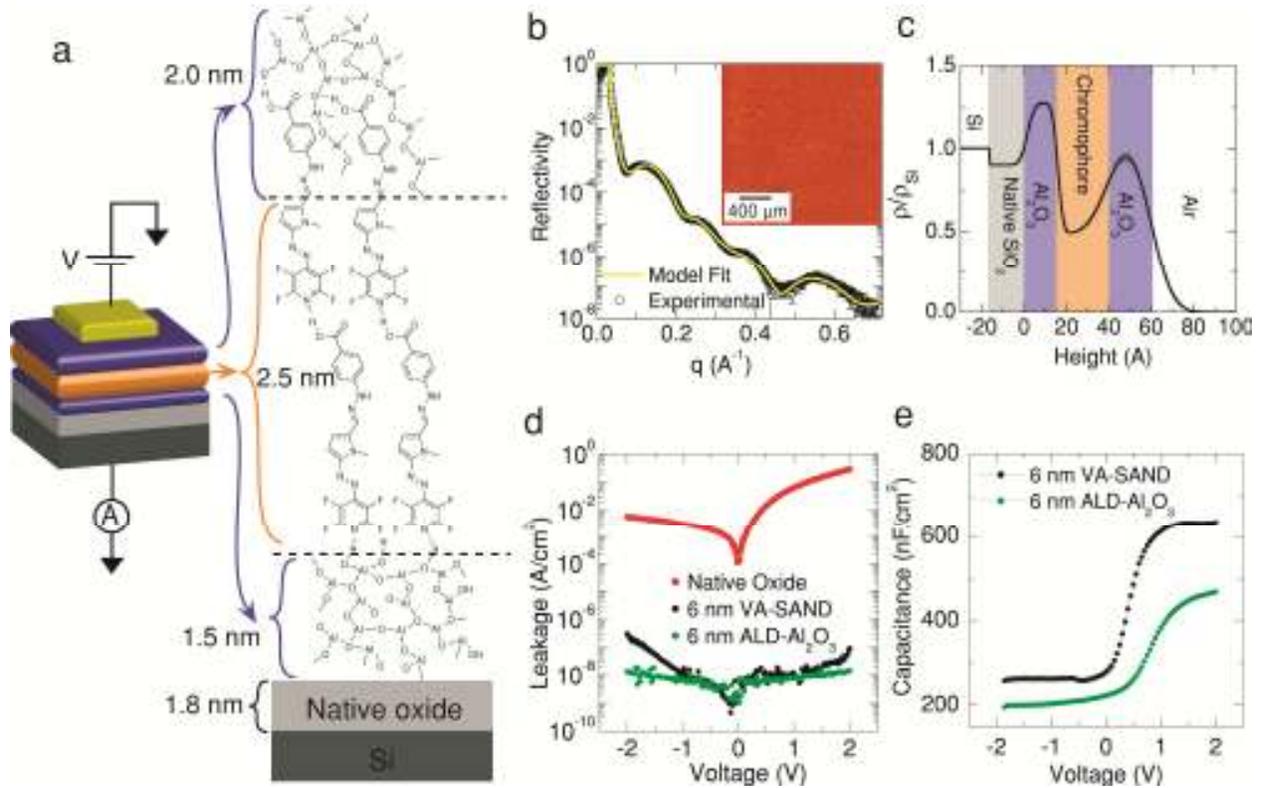

**Figure 1.** Structure and properties of VA-SAND gate dielectrics. a) TFT schematic and chemical structure of VA-SAND on Si/SiO$_2$ substrates. b) X-ray reflectivity data and best-fit results plotted as a function of the momentum transfer vector ($q = 4\pi \sin(2\theta/2)/\lambda$, where $2\theta$ = angle of the scattered X-rays and $\lambda$ = X-ray wavelength). Inset shows an AFM image of the VA-SAND surface with RMS roughness = 0.65 nm. c) Extracted electron density profile of VA-SAND, corresponding to best-fit results noted in a), as a function of height (Z) from the native-oxide surface showing the densities of constituent layers. d) Leakage current density of MIS fabricated on VA-SAND compared to that of 6 nm Al$_2$O$_3$ and native oxide on Si. e) Capacitance of VA-SAND and 6 nm Al$_2$O$_3$ as a function of top-electrode voltage at 10 kHz.

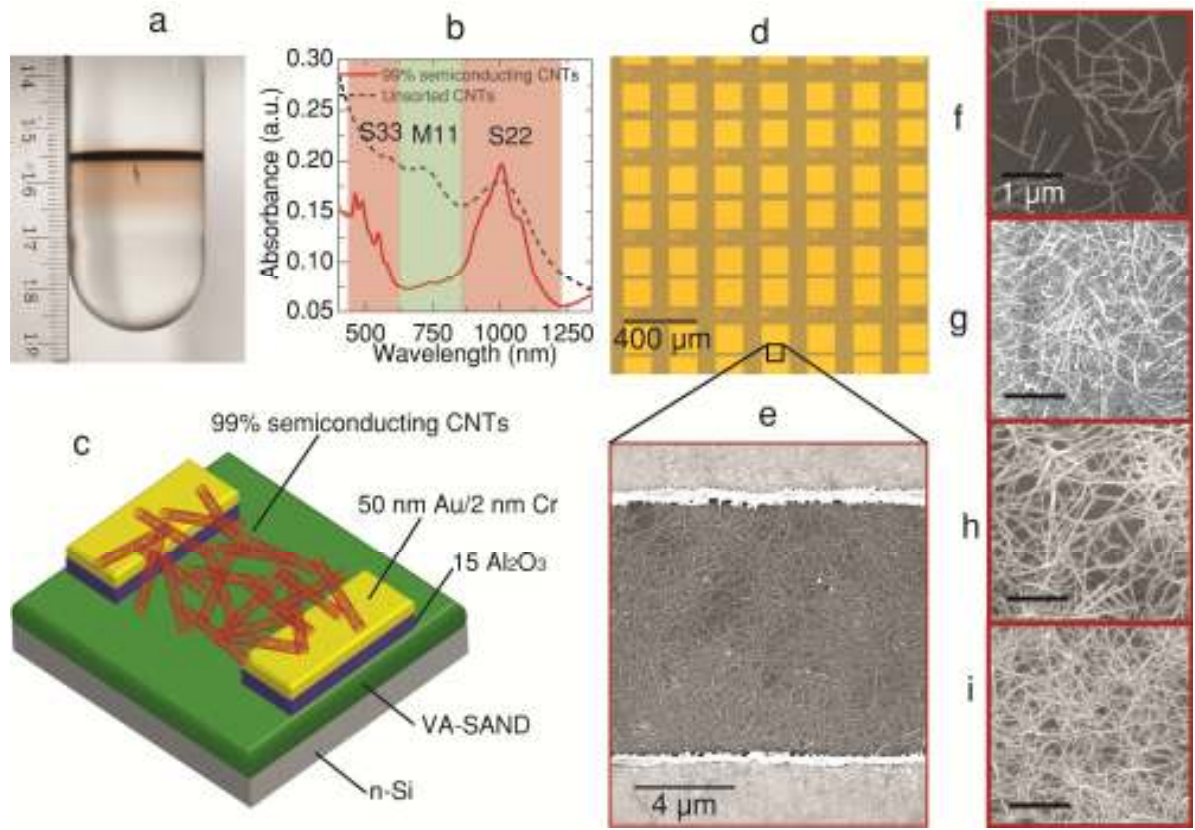

**Figure 2.** Architecture and channel morphology of CNT TFTs. a) Optical micrograph of a centrifuge tube containing a 99% semiconducting CNT band after two iterations of density gradient ultracentrifugation. b) Optical absorbance of sorted semiconducting CNTs compared with that of diluted unsorted CNTs to highlight semiconducting purity. Due to the different concentrations of CNTs in each solution, the absolute peak heights cannot be directly compared. c) Schematic of a bottom-contact random CNT TFT fabricated on VA-SAND. d) Optical micrograph of a large array of CNT TFTs with varying channel lengths. e) Scanning electron microscopy (SEM) image of a CNT channel. f)-i) SEM images of CNT thin-films with density-1, density-2, density-3, and density-4, respectively, as discussed in the text. Scale bars in Figs. 2f-i correspond to 1 µm.

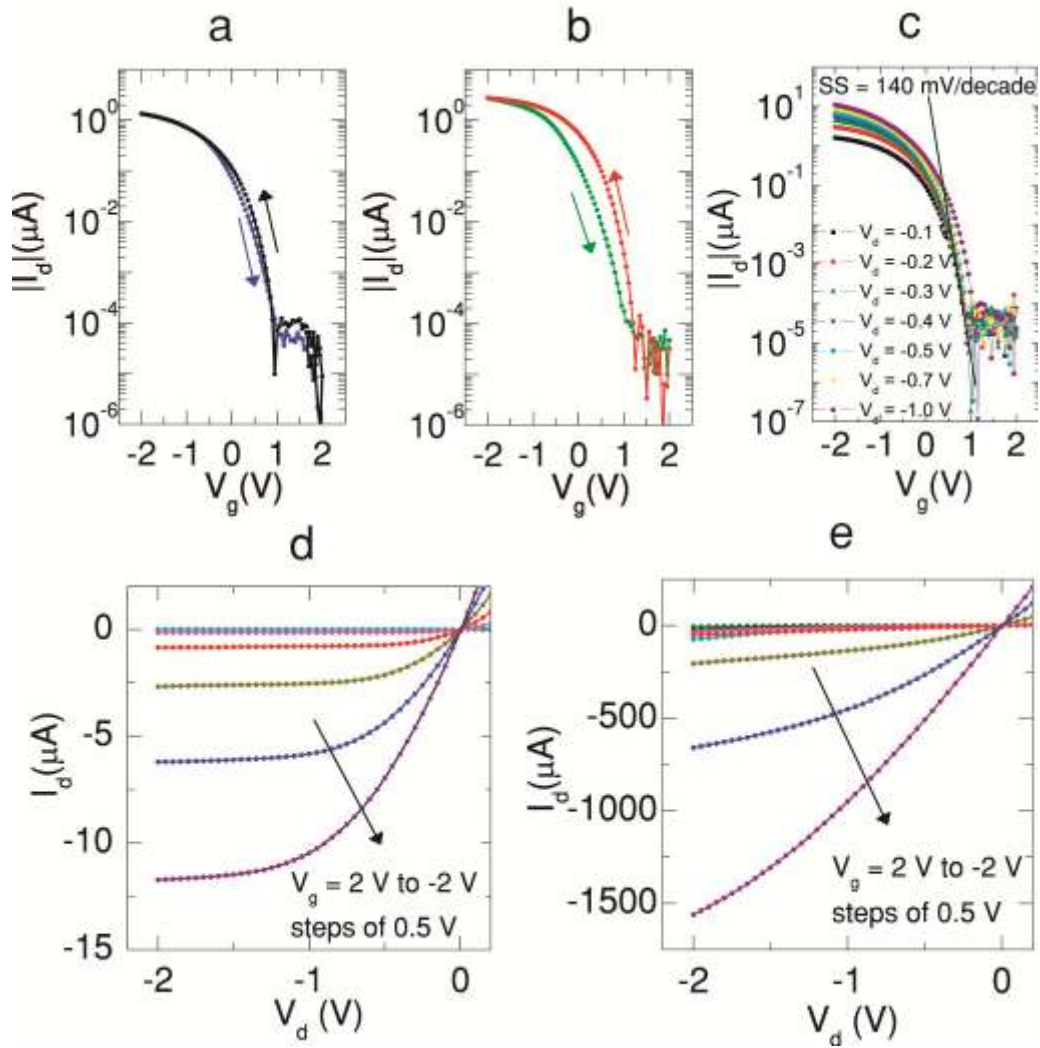

**Figure 3.** Transfer and output characteristics of CNT TFTs. a) Transfer characteristics of a density-1 CNT TFT ($L = 5$ μm, $W = 100$ μm) on VA-SAND with forward and backward sweeps showing negligible hysteresis. b) A density-1 CNT TFTs ($L = 5$ μm, $W = 100$ μm) on 6 nm $Al_2O_3$ showing increased hysteresis. c) Transfer characteristics of the same device as in (a) on VA-SAND showing low sub-threshold slope (140 mV/decade) for drain bias ($V_d$) varying from -0.1 V to -1 V. d) and e) Output characteristics of CNT TFTs ($L = 5$ μm, $W = 100$ μm) with lowest CNT density (5.5 CNTs/μm$^2$) and highest CNT density (27.1 CNTs/μm$^2$).

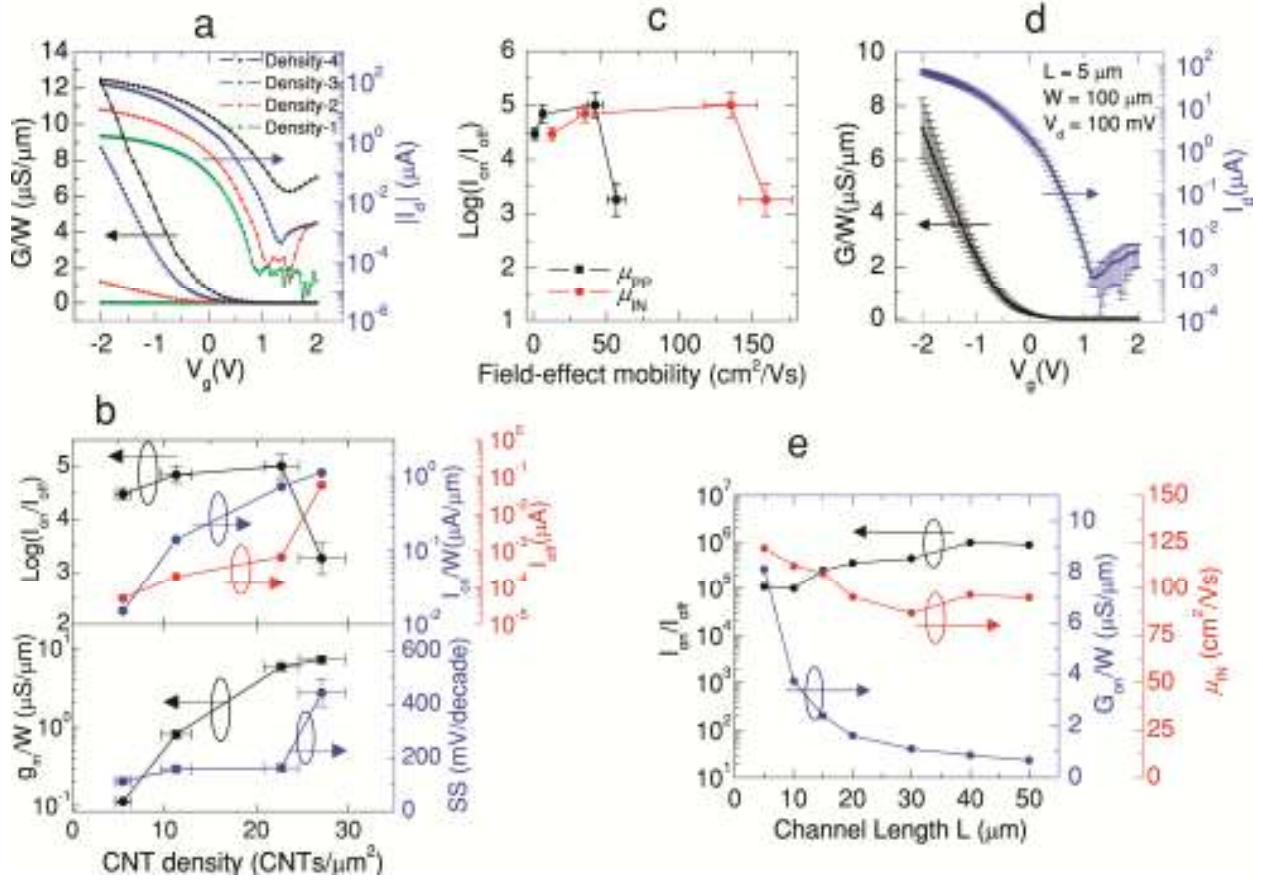

**Figure 4.** CNT density and channel geometry dependent characteristics of CNT/VA-SAND TFTs. a) Transfer characteristics showing width-normalized conductance ($G/W = I_d/(V_d \cdot W)$) and drain current ($I_d$) of 4 CNT TFTs ($L = 5$ μm, $W = 100$ μm) with CNT density varying from density-1 (5.5 CNTs/μm$^2$) to density-4 (27.1 CNTs/μm$^2$). b) Average device parameter normalized on-current $I_{on}/W$, off-current $I_{off}$, log of on/off ratio ($\text{Log}(I_{on}/I_{off})$), normalized transconductance $g_{m,nor}$ and sub-threshold slope $SS$ plotted as a function of CNT density for CNT TFTs from Fig. 4a. Horizontal and vertical error bars represent stand deviation in CNT density and exponent $m$ of on/off ratio ($10^m$), respectively. c) $\text{Log}(I_{on}/I_{off})$ plotted as a function of field-effect mobility (see text) for 4 different CNT densities. $\text{Log}(I_{on}/I_{off})$ and field-effect mobility are averaged over 5 devices. d) Average transfer characteristics of 7 density-3 (22.7 CNTs/μm$^2$)

CNT TFTs ($L$ = 5 μm, $W$ = 100 μm). e) Transfer curves and f) device parameters, on/off ratio, normalized on-state conductance $G_{on}/W$ and intrinsic field-effect mobility $\mu_{IN}$ of density-3 CNT TFTs as a function of channel length $L$ varying from $L$ = 5-50 μm (W = 100 μm).

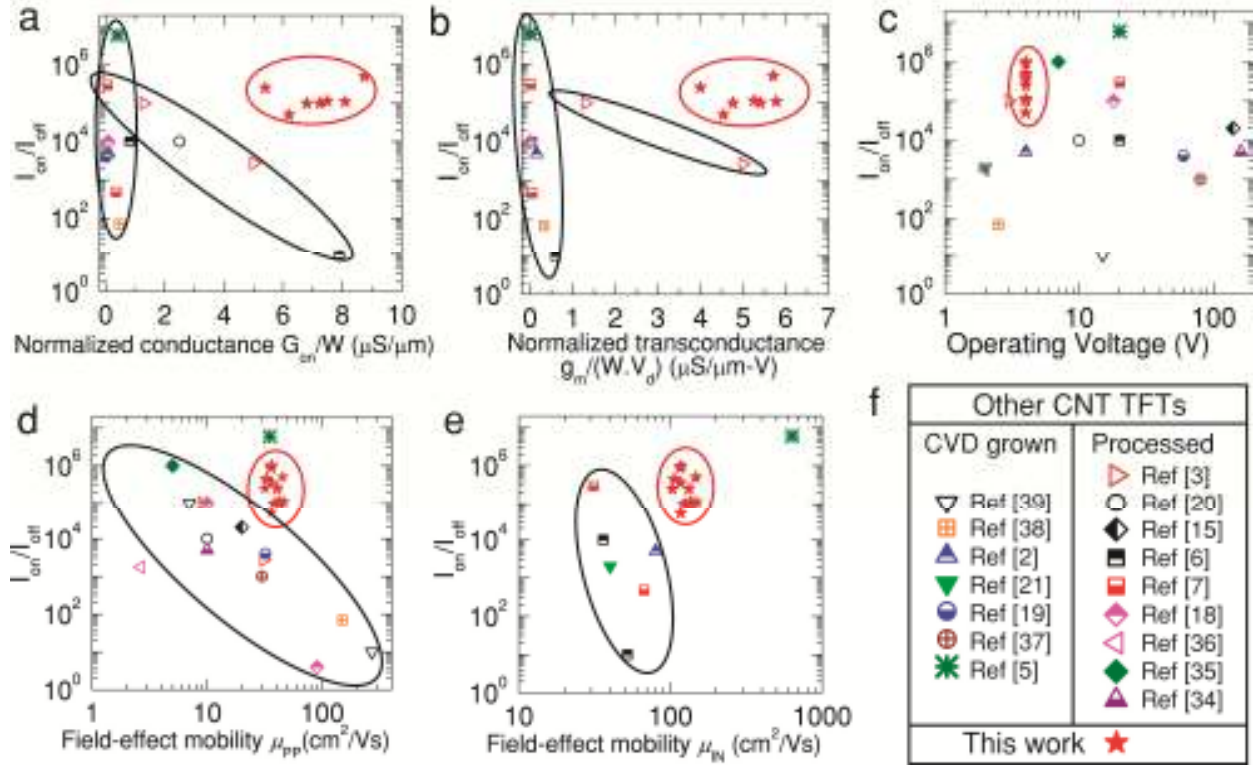

**Figure 5.** Comparison of CNT/VA-SAND TFT response with literature precedent. On/off ratio *versus*: a) On-state normalized conductance $G_{on}/W$, b) Normalized transconductance $g_m/(W \cdot V_d)$, c) Operating voltage of density-3 (22.7 CNTs/µm²) CNT TFTs on VA-SAND ($L$ = 5 µm, $W$ = 100 µm), compared with device design trade-off trends for previously reported CNT TFTs. Plots of on/off ratio *versus* d) Parallel plate field-effect mobility $\mu_{PP}$, and e) Intrinsic field-effect mobility $\mu_{IN}$ of density-3 CNT TFTs ($W$ = 100 µm) on VA-SAND are compared with previously reported CNT TFTs. f) Legend for previously reported CNT TFTs for all plots, Figs. 5a-e.

## ASSOCIATED CONTENT

**Supporting Information.** Density gradient ultracentrifugation of 99% semiconducting single-walled carbon nanotubes, length distribution of monodisperse CNTs, calculations of field-effect mobility are provided. This material is available free of charge *via* the Internet at http://pubs.acs.org.

## AUTHOR INFORMATION

**Corresponding Authors**

*Address: Northwestern University, Department of Materials Science and Engineering, 2220 Campus Drive, Evanston, IL 60208. Tel: (847) 491-2696, Email: m-hersam@northwestern.edu.

*Address: Northwestern University, Department of Chemistry, 2145 Sheridan Road, Evanston, IL 60208. Tel: (847) 491-5658, Email: t-marks@northwestern.edu.

**Author Contributions**

M.C.H, T.J.M., L.J.L., and V.K.S. conceived the experiments, analyzed and interpreted data. R.P.O., J.M.P.A. and V.K.S. fabricated VA-SAND gate dielectrics. J.D.E. and M.J.B. conducted X-ray reflectivity experiments and analyzed data. V.K.S. fabricated the devices, and conducted the measurements. All authors contributed to the discussion and writing of the manuscript.

## ACKNOWLEDGMENTS

This research was supported by the National Science Foundation (DMR-1006391 and DMR-1121262) and by the Nanoelectronics Research Initiative at the Materials Research Center of Northwestern University. The use of the J.B. Cohen X-Ray Diffraction Facility was supported through the MRSEC program at the Materials Research Center of Northwestern University. R.

P. O. acknowledges funding from the European Community's Seventh Framework Programme through a Marie Curie International Outgoing Fellowship (Grant Agreement 234808). We thank Drs. R. Divan and L. Ocala of the Center for Nanoscale Materials, Argonne National Laboratory, for assistance with clean room fabrication.

**Table of Contents Figure**

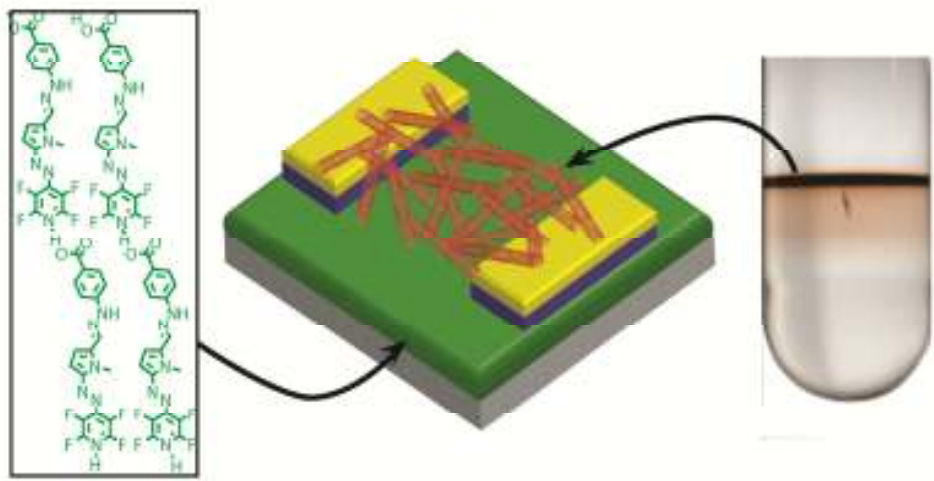